\documentclass[12pt]{iopart}
\usepackage{iopams,amssymb,graphicx,dsfont}
\newcommand{\openone}{\mathbf{1}}
\newcommand{\eqref}[1]{(\ref{#1})}
\begin{document}
\title[Random-matrix theory of  resonators with  $\mathcal{PT}$ or $\mathcal{PTT}'$
symmetry]{Random-matrix theory of
amplifying and absorbing resonators with $\mathcal{PT}$ or $\mathcal{PTT}'$
symmetry}
\author{Christopher Birchall and Henning Schomerus}
\address{Department of Physics, Lancaster University, Lancaster, LA1 4YB, United Kingdom}
\date{\today}
\begin{abstract}
We formulate gaussian and circular random-matrix models representing a
coupled system consisting of an absorbing and an amplifying resonator,
which are mutually related by a generalized time-reversal symmetry.
Motivated by optical realizations of such systems we consider a
$\mathcal{PT}$ or a $\mathcal{PTT}'$ time-reversal symmetry, which impose
different constraints  on magneto-optical effects, and  then focus on
five common settings. For each of these, we determine the eigenvalue
distribution in the complex plane in the short-wavelength limit, which
reveals that the fraction of real eigenvalues among all eigenvalues in the spectrum  vanishes if all
classical scales are kept fixed. Numerically, we find that the transition
from real to complex eigenvalues  in the  various ensembles display a
different  dependence on the coupling strength between the two
resonators. These differences can be linked to the level spacing
statistics in the hermitian limit of the considered models.
\end{abstract}

\submitto{\JPA}
\maketitle

\section{Introduction}

The investigation of nonhermitian $\mathcal{PT}$-symmetric Hamiltonians
is motivated by the fact that they possess eigenvalues which  are either
real or occur in complex-conjugate pairs \cite{bender}. Considerable
attention has been paid to the delineation of systems with a completely
real spectrum, with many works focussing on exactly solvable
one-dimensional situations (for reviews see \cite{ptreview1,ptreview2}).
With the recent advent of optical implementations
\cite{ptexperiments1,ptexperiments2} it has been realized that the
appearance of complex eigenvalues drives a number of interesting
switching effects
\cite{ptexperiments1,ptexperiments2,christodoulidesgroup1,christodoulidesgroup2,christodoulidesgroup3,
pteffects1,pteffects2a,pteffects4,pteffects5}, including the possible
onset of lasing
\cite{hsqoptics,longhilaserabsorber,variousstonepapers1,yoo}, which moves
the most unstable states (with energies or frequencies that have a large
positive imaginary part) into the centre of attention. At the same time,
these implementations motivate the study of multi-dimensional systems in
which many modes become mixed by multiple scattering. Here, we
investigate the formation and distribution of the complex spectrum in
such situations on the basis of a statistical approach rooted in
random-matrix theory \cite{mehta,haake}, which samples systems that
respect a certain set of symmetries and share a number of well-defined
characteristic energy and time scales, but differ in the microscopic
details of the dynamics.

We extend  earlier exploratory works of this approach
\cite{hsrmt,birchallweyl} to consider random-matrix ensembles which
differ by the assumed absence or presence of elastic or dissipative
magneto-optical effects. This leads to a choice between two generalized
time-reversal symmetries, termed $\mathcal{PT}$ and $\mathcal{PTT}'$
symmetry and physically motivated in \cite{hsreview}. These ensembles
apply to a coupled-resonator geometry (with an absorbing resonator
possessing $M$ internal modes coupled to a matching amplifying resonator
via an interface of $N$ channels with transparency $T$, and amplification
or absorption rate set to a common value $\mu$). The optical setting
motivates to consider 5 particular scenarios (OO, UO, UO$'$, OA and
OA$'$). These can be studied either based on an effective Hamiltonian or
in terms on an effective time-evolution operator (a quantum map), as is
described in section \ref{sec:ensembles}.

In section \ref{sec:largem} we determine for each scenario the
distribution of eigenvalues in the complex plane in the short-wavelength
limit $M\to\infty$ at fixed $\alpha=M/N$, $T$ and $\mu$. We find that in this limit, the
fraction of real eigenvalues among all eigenvalues vanishes at any finite fixed amplification and
absorption rate, with the details of the eigenvalue distribution in the
complex plane depending on the symmetry class. This supports the
conclusion of earlier work on some of these ensembles \cite{hsrmt} that
the transition to the complex spectrum occurs at a characteristic
absorption rate $\mu_\mathrm{PT}$ which is classically small when
compared to the inverse dwell time
$E_\mathrm{T}=1/t_\mathrm{dwell}=NT\Delta/2\pi$ in each part of the
resonator, but large when compared to the mean level spacing $\Delta$.

In order to investigate the details of this transition we then present
results of extensive numerical investigations (section
\ref{sec:transition} and \ref{sec:spacings}). These reveal that the
various ensembles display a different dependence of the transition on the
coupling strength $T$, as well as on $M$ and $N$. This
division is associated with specific mechanisms of eigenvalue
coalescence, which we relate to differences in the level spacing
statistics in the hermitian limit by extending the
perturbative considerations of \cite{hsrmt}.

Section \ref{sec:conclusions} contains our conclusions.

Throughout this work we denote eigenvalues as $E$, but set $\hbar\equiv
1$; all considerations thus directly apply to the eigenfrequencies in
optical analogues of non-hermitian quantum systems.

\section{\label{sec:ensembles}Random-matrix ensembles}

\begin{figure}
\center{\includegraphics[width=0.6\textwidth]{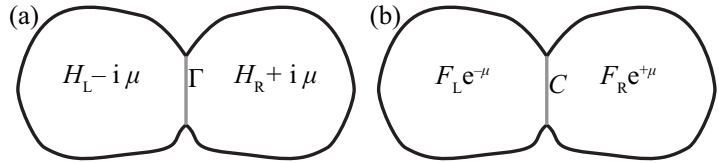}}
\caption{\label{fig1}Illustration of (a) the effective Hamiltonian \eqref{eq:h} and (b) the quantum map  \eqref{eq:map} used to model an absorbing resonator
(L) which is symmetrically coupled (via an interface characterized by $\Gamma$ or $C$) to an amplifying resonator (R). Various
symmetry classes arise depending on the  constraints imposed on the internal Hamiltonians $H_\mathrm{L}$ and $H_\mathrm{R}$.
The depicted situation applies to uniform amplification or absorption with rate $\mu$;
further symmetry classes arise when $\mu$ is replaced by matrices $X_\mathrm{L}$ and $X_\mathrm{R}$.
}
\end{figure}

Following \cite{hsrmt,birchallweyl,hsreview}, we consider a
coupled-resonator geometry where one part of the system (L) is absorbing
and the other part (R) is amplifying, with the absorption and
amplification rate set to a matching value $\mu$. In each part random
multiple scattering at a rate $1/\tau$ results in a mixing of
$M=1/\Delta\tau$ internal modes, where $\Delta$ is the mean level
spacing, and the two parts are coupled together at an interface which
supports $N$ open channels of transparency $T_n$, $n=1,2,\ldots,N$. In
order to capture the consequences of multiple scattering we utilize
effective Hamiltonians and quantum maps, which model these systems as
illustrated in figure \ref{fig1}.

\subsection{Effective Hamiltonians and symmetry classes}

The general structure of an effective Hamiltonian for this situation can
been derived in a systematic scattering approach \cite{hsrmt,hsreview}.
This yields
\begin{equation}\label{eq:h}
\mathcal{H}=
\left(\begin{array}{cc}H_\mathrm{L}-\rmi X_\mathrm{L} & \Gamma \\  \Gamma & H_\mathrm{R} +\rmi X_\mathrm{R}\\  \end{array} \right),
\end{equation}
where the $M\times M$-dimensional hermitian matrices $H_\mathrm{L}$ and
$H_\mathrm{R}$ ($X_\mathrm{L}$ and $X_\mathrm{R}$) represent the internal
hermitian (anti-hermitian) dynamics in each part of the system, while the
coupling matrix is of the specific form
\begin{equation}\label{eq:gamma}
\Gamma=\frac{\Delta M}{\pi}\mathrm{diag}(\underbrace{\gamma_1,\ldots,\gamma_N}_{\mbox{\scriptsize $N$ entries}},\underbrace{0,\ldots,0}_{\mbox{\scriptsize $M-N$ entries}}),\quad
\gamma_n=\frac{\sqrt{T_n}}{1+\sqrt{1-T_n}}.
\end{equation}
The specific form of \eqref{eq:h} displays a structure which goes beyond
the mere symmetry requirements usually applied in mathematical
classifications of nonhermitian matrices (for comprehensive overviews see
\cite{magnea,akemann1,akemann2}). In particular, the matrix $\Gamma$
needs to be positive definite and bounded in order to model physical
coupling between two resonators. This structure resembles analogous
models in mesoscopic superconductivity, where the two subspaces represent
particles and holes, and the coupling is provided by Andreev reflection
\cite{melsen,zirnbauer,altlandzirnbauer}.

We now impose two different versions of generalized time-reversal
symmetry. Traditional $\mathcal{PT}$ symmetry involves the parity
operator ${\cal P}=\sigma_x\otimes \openone_M$, where the Pauli matrix
$\sigma_x$ interchanges the subspaces R and L, as well as the time
reversal operation ${\cal T}=K$, where $K$ is the complex conjugation in
a given basis, assumed to coincide with the basis  of \eqref{eq:h}.
Invariance under the joint $\mathcal{PT}$  operation then demands
\begin{equation}
\mathcal{H}={\cal P}{\cal H}^*{\cal P}  \quad \Rightarrow \quad H_\mathrm{L}=H_\mathrm{R}^*=H_\mathrm{R}^T, \quad X_\mathrm{L}=X_\mathrm{R}^*=X_\mathrm{R}^T.
\end{equation}
In a $\mathcal{PT}$-symmetric basis, the secular polynomial
$s(E)=\mathrm{det}\,(\mathcal{H}-E\openone_{2M})$ has real coefficients,
which constraints each eigenvalue to be real or being partnered by its
complex conjugate.

In hermitian situations, the complex conjugation $\mathcal{T}$ is
equivalent to taking the transpose of the matrix (thus passing from the
right eigenvalue problem to the left eigenvalue problem). In
non-hermitian situations, this transposition amounts to an  independent
operation, denoted as $\mathcal{T}'$ \cite{hsrmt,hsreview}. For a
$\mathcal{PTT}'$-symmetric situation, we now obtain the constraints
\begin{equation}
\mathcal{H}={\cal P}{\cal H}^\dagger{\cal P}  \quad \Rightarrow \quad H_\mathrm{L}=H_\mathrm{R}, \quad X_\mathrm{L}=X_\mathrm{R},
\end{equation}
which is of interest as this yields the same spectral constraints as
$\mathcal{PT}$ symmetry.

For each of these two cases, a number of ensembles can now be formulated
depending on the presence or absence of additional symmetries for
$H\equiv H_\mathrm{L}$ and $X\equiv X_\mathrm{L}$. In particular, we
consider the cases that they may be further constrained to be real and
thus symmetric (labeled O for orthogonal), complex (labeled U for
unitary), or purely imaginary and thus antisymmetric (labeled A). In
combination, we then arrive at 9 symmetry classes with
$\mathcal{PT}$-symmetry, denoted as $S_\mathrm{H}S_\mathrm{X}$,
$S_i=\mathrm{O},\mathrm{U},\mathrm{A}$, as well as 8 additional classes
$S_\mathrm{H}S_\mathrm{X}'$ with $\mathcal{PTT}'$-symmetry (OO and OO$'$
coincide as in this case $\mathcal{T}'$ is an independent symmetry).

A detailed discussion of the physical requirements corresponding to the
various symmetries in optical settings is given in \cite{hsreview}.
Motivated by this context, we focus on 5 situations, OO, UO, UO$'$, OA
and OA$'$. The most important scenario is that of OO symmetry, with
$H=H^T=H^*$, $X=X^T=X^*$, which includes ordinary optical systems with
gain and loss modeled by a complex refractive index. The cases of UO and
UO$'$ symmetry ($X=X^T=X^*$ but $H$ not further constrained) model
systems with elastic magneto-optical effects, with different symmetry
constraints imposed on the effective magnetic field. We will also
consider the OA and OA$'$ cases, as it is known that absorption can be
provided by magneto-optical devices \cite{stoffregen} (in practice,
however the design of a matching magneto-optical amplification may prove
challenging).

\subsection{Random-matrix ensembles}

The described symmetry classes are converted into statistical ensembles
under convenient sampling  of the $M\times M$-dimensional hermitian
matrices $H$ and $X$. Specifically, depending on whether
$S_\mathrm{H}=\mathrm{O}$ or U we choose $H$ from the standard Gaussian
orthogonal or unitary ensemble (GOE or GUE) of random matrix theory
\cite{mehta,haake}, respectively. The variance $|H_{lm}|^2=\sigma=1/M$ of
the matrix elements is set such that the probability distribution of
eigenvalues $E$ becomes stationary in the large-$M$ limit, corresponding
to a Wigner semicircle law with radius 2,
 \begin{equation}\label{eq:semicircle}
\bar\rho(E)=\pi^{-1}\sqrt{1-E^2/4}.
\end{equation}

For $S_\mathrm{X}=\mathrm{O}$ symmetry of the anti-hermitian part, we
model uniform absorption and amplification by setting $X=\mu\openone_M$.
For the case $S_\mathrm{X}=\mathrm{A}$, we model $\rmi X=-A$ via a real
antisymmetric matrix with random Gaussian elements, and quantify the
degree of non-hermiticity by $\mu^2=M^{-1}\mathrm{tr} \overline{AA^T}$.

Throughout, we will model the coupling between the two parts of the
system via $N\equiv \alpha M$ channels of identical transparency $T$.
Together with the chosen energy scaling \eqref{eq:semicircle}, which
gives $\Delta_0\equiv M/\rho(E=0)=M/\pi$, the coupling matrix
\eqref{eq:gamma} then takes the form
$\Gamma=\mathrm{diag}(\gamma,\ldots,\gamma,0,\ldots,0)=\gamma\,\mathrm{diag}(\openone_{N},0_{M-N})$,
with $N$ finite diagonal entries $\gamma=\sqrt{T}(1+\sqrt{1-T})^{-1}$.

We denote these ensembles as G$S_\mathrm{H}S_\mathrm{X}$E or
G$S_\mathrm{H}S_\mathrm{X}$E$'$, and specifically consider the cases
GOOE, GUOE, GUOE$'$, GAOE, and GAOE$'$, which correspond to the optical
settings described in the previous subsection.

\subsection{Effective quantum maps}
An alternative approach in random-matrix theory  bases the considerations
on circular ensembles of effective time-evolution operators
\cite{mehta,haake}. For the coupled-resonator geometry, the general
structure of these operators has been identified in
\cite{birchallweyl,hsreview}. They take the form of a quantum map
\begin{equation}\label{eq:map}
\fl
{\cal F}=
\sqrt{C}
\left(\begin{array}{cc} \rme^{-\mu\tau}F_\mathrm{L} & 0 \\ 0 &  \rme^{\mu\tau}F_\mathrm{R} \\ \end{array}  \right)
\sqrt{C},
\quad \sqrt{C}=
\left(\begin{array}{cc} \mathrm{Re}\,\tilde\gamma\,P+Q & -\rmi\mathrm{Im}\,\tilde\gamma\,P\\
-\rmi\mathrm{Im}\,\tilde\gamma\,P  &\mathrm{Re}\,\tilde\gamma\,P+Q \\ \end{array}  \right)
,
\end{equation}
which delivers quasienergies $E_n$ via the eigenvalue problem
\begin{equation}
{\cal F}\psi_n=\lambda_n\psi_n,\quad
\lambda_n=\exp(-\rmi E_n\tau).
\end{equation}
The properties of the interface are now encoded in the parameter
$\tilde\gamma=\sqrt{\sqrt{R}+\rmi\sqrt{T}}$, the rank-$N$ projector
$P=\mathrm{diag}(\openone_{N},0_{M-N})$, and the complementary projector
$Q=\openone_M-P$. The internal dynamics are described by the $M\times
M$-dimensional unitary matrices $F_\mathrm{L}$ and $F_\mathrm{R}$, which
satisfy $F_\mathrm{L}=F_\mathrm{R}^T$ for $\mathcal{PT}$ symmetry, and
$F_\mathrm{L}=F_\mathrm{R}$ for $\mathcal{PTT}'$ symmetry. Finite $\mu$
breaks the unitarity of the quantum map.  (In the specified form,
\eqref{eq:map} holds for $S_\mathrm{X}=\mathrm{O}$ symmetry with uniform
amplification and absorption, but by the replacement $\mu\to X$ can be
adapted to other symmetries.) Appropriate random-matrix ensembles follow
by choosing $F=F_\mathrm{L}$ from the standard circular orthogonal or
unitary ensembles (COE or CUE), respectively \cite{mehta,haake}.  We
denote these circular ensembles with $\mathcal{PT}$ and $\mathcal{PTT}'$
symmetry as C$S_\mathrm{H}S_\mathrm{X}$E and
C$S_\mathrm{H}S_\mathrm{X}$E$'$, respectively.

\subsection{Overview of characteristic parameters and scales}
In summary, each RMT ensemble is  specified by the symmetry
$S_\mathrm{H}S_\mathrm{X}$, as well the following 4 dimensionless
parameters:  the number of modes $M=1/\Delta\tau$ in each of the two
parts of the system, the relative size $\alpha=N/M$ of the interface, the
transparency $T$ of the interface (encoded in $\gamma$ or $\tilde
\gamma$), and the degree of non-hermiticity $\mu/E_\mathrm{T}$, where
$E_\mathrm{T}\equiv NT/2\pi\Delta$ is the Thouless parameter mentioned in
the introduction. In the Hamiltonian variants of RMT, $\Delta=\pi/M$ and
$E_\mathrm{T}=T\alpha/2$, while in the quantum map version with
$\tau\equiv 1$, $\Delta=2\pi/M$ and $E_\mathrm{T}=T\alpha$.

\section{\label{sec:largem}Eigenvalue distribution in the large $M$ limit}

In order to get insight into the distribution of eigenvalues in the
complex plane, and the conditions under which they may accumulate on the
real axis, we first consider  the limit of a large number of internal
modes $M\to \infty$, at fixed $\alpha=N/M$, $\mu/E_\mathrm{T}$ and $T$.
For an optical system, this limit is realized by decreasing the
wavelength (increasing the frequency) in a given resonator geometry while
keeping the absorption and amplification rate $\mu$ at a
wavelength-independent value. In random-matrix theory, this limit can be
approached via systematic diagrammatic expansions,
where the leading order captures the averaged eigenvalues density
neglecting fluctuations on the scale of the level spacing
\cite{melsen,janik1,janik2,fyodorovsommers2}. We now adapt this approach to the
symmetries in question.

\subsection{Generalized Pastur equation}

The effective Hamiltonian ${\cal H}$ generally possesses complex
eigenvalues, and the complex-analysis nature of the method employed below
suggests to denote these as $z$. The distribution of eigenvalues in the
complex plane can then be written as
\begin{equation}
\rho(z,z^*)=\frac{1}{2M}\frac{1}{\pi}\frac{\partial{{\rm tr}\,{\cal
G}_{11}}}{\partial z^*},
\end{equation} where ${\cal
G}_{11}$ denotes the $2M\times 2M$-dimensional  top-left block of the
$4M\times 4M$-dimensional matrix Green function
\begin{equation}
{\cal G}=\left(
           \begin{array}{cc}
             z-{\cal H} & \rmi\lambda \\
             \rmi\lambda & z^*-{\cal H}^\dagger\\
           \end{array}
         \right)^{-1}.
\end{equation}
The limit $\lambda\to 0$ is implied to be taken at the end of the
calculation.

In order to work out the random-matrix average we expand the matrix Green
function as a geometric series
\begin{eqnarray}
{\cal G}&=& {\cal U}^{-1}\sum_{n=0}^\infty(-{\cal H}_0{\cal U}^{-1})^n, \quad {\cal H}_0=
{\rm diag}(H,H,H,H) ,
\\
{\cal U}&=&\left(
           \begin{array}{cccc}
             z+\rmi\mu & -\Gamma &   \rmi\lambda  & 0\\
              -\Gamma & z-\rmi\mu    & 0 & \rmi\lambda\\
              \rmi\lambda  & 0   & z^*-\rmi\mu & -\Gamma    \\
              0 & \rmi\lambda &  -\Gamma & z^*+\rmi\mu    \\
           \end{array}
         \right)^{-1},
         \label{eq:geom}
\end{eqnarray}
where the momentarily stipulated form of ${\cal H}_0$ holds for the GOOE
and GUOE$'$ ensembles (the other ensembles are discussed thereafter).

The average can now be carried out by contractions of the Gaussian random
variables in $H$, which can be represented diagrammatically. The leading
order (the planar limit) is given by rainbow diagrams in which the contraction lines do not
cross,
\begin{equation}
\bar{\cal G}={\cal U}^{-1}+{\cal U}^{-1} \mathop{\mathcal{H}_0\bar{\cal G}\mathcal{H}_0}_{
\raise 1ex \hbox{\vrule width .1ex height 0pt depth 1ex}%
\lower 0ex\hbox{\vrule width 4.5ex height 0pt depth .1ex}%
\raise 1ex \hbox{\vrule width .1ex height 0pt depth 1ex}%
}\bar{\cal G},
\end{equation}
which sum up to
\begin{equation}
\bar{\cal G}={\cal U}^{-1}+{\cal U}^{-1}[G\otimes \openone_M]\bar{\cal G}
\Rightarrow
\bar{\cal G}=({\cal U}^{-1}-G\otimes \openone_M)^{-1},
\label{eq:rainbow}
\end{equation}
where $G=\frac{1}{M}{\rm tr}_M\,\bar {\cal G}$ is a reduced $4\times 4$
matrix Green function.

The matrix ${\cal U}^{-1}-G\otimes \openone_M$ on the right-hand side of
\eqref{eq:rainbow} separates into $N$ blocks of the form $(u_\gamma-G)$
and $M-N$ blocks of the form $(u_0-G)$, where
\begin{equation}
u_\gamma=\left(
                 \begin{array}{cccc}
                   z+\rmi\mu & -\gamma & \rmi\lambda & 0 \\
                   -\gamma & z-\rmi\mu & 0 & \rmi\lambda \\
                   \rmi\lambda & 0 & z^*-\rmi\mu & -\gamma \\
                   0 & \rmi\lambda & -\gamma & z^*+\rmi\mu  \\
                 \end{array}
               \right).
\end{equation}
We thus can invert each block separately, and take the partial trace on
both sides. This leads to the generalized Pastur equation
\begin{equation}
G=\alpha (u_\gamma- G)^{-1}+(1-\alpha)  (u_0- G)^{-1}
\label{eq:pastur}
\end{equation}
for the GOOE and GUOE$'$ ensembles.

For the  GUOE ensemble, \eqref{eq:geom} holds with ${\cal H}_0= {\rm
diag}(H,H^*,H,H^*)={\rm diag}(H,H^T,H,H^T)$. The transpositions reduce
the number of rainbow diagrams in the Gaussian average, which leads to
the modified equation
\begin{equation}
G=\alpha (u_\gamma- P_1 G P_1-P_2 G P_2)^{-1}+(1-\alpha)  (u_0- P_1 G P_1-P_2 G P_2)^{-1},
\label{eq:pastur2}
\end{equation}
where  $P_1={\rm diag}\,(1,0,1,0)$ and $P_2={\rm diag}\,(0,1,0,1)$.

For the  GOAE ensemble, the random matrix $A$ has to be incorporated into
${\cal H}_0= {\rm diag}(H+A,H+A,H-A,H-A)$, while $\mathcal{U}$ is
replaced by $\tilde\mathcal{U}=\mathcal{U}|_{\mu=0}$. Instead, $\mu$
appears via the contractions of $A$. This leads to the condition
\begin{equation}
G=\alpha (\tilde u_\gamma- G +RGR)^{-1}+(1-\alpha)  (\tilde u_0- G +RGR)^{-1},
\label{eq:pastur3}
\end{equation}
where $R=\mu\,\mathrm{diag}(1,1,-1,-1)$ and $\tilde u_\gamma=u_\gamma|_{\mu=0}$.

Finally, in the GOAE$'$ ensemble we have ${\cal H}_0=
{\rm diag}(H+A,H-A,H-A,H+A)$, and
\begin{equation}
G=\alpha (\tilde u_\gamma- G +R'GR')^{-1}+(1-\alpha)  (\tilde u_0- G +R'GR')^{-1},
\label{eq:pastur4}
\end{equation}
where $R'=\mu\,\mathrm{diag}(1,-1,-1,1)$.

\subsection{Solution of the generalized Pastur equation}

The condition \eqref{eq:pastur} can be rephrased as
\begin{equation}
(u_\gamma- G)G(u_0- G)=  u_{(1-\alpha)\gamma}-G ,
\end{equation}
thus, a third-degree matrix polynomial, and the versions
\eqref{eq:pastur2}--\eqref{eq:pastur4} can be rewritten analogously. If
we write $G$ in terms of its 16 components and eliminate these
successively, we end up with a polynomial of a very large degree, which
prohibits an exact analytical solution. Therefore,  we pursue a
semi-analytical approach which starts with the exact solution $G_0$ in
the uncoupled case $\alpha=0$, where $G_0(u_0- G_0)=  \openone_4$
(independently of $\gamma$). This is solved by
$G_0=u_0/2+(u_0^2/4-\openone_4)^{1/2}$, where the square root of the
matrix $K=u_0^2/4-\openone_4=V{\rm diag}\,{k_n}V^{-1}$ is defined by
diagonalization, $K^{1/2}=V{\rm diag}\,(\pm\sqrt{k_n})V^{-1}$, which here
can be carried out explicitly as $K$ decouples into two independent
$2\times 2$-dimensional blocks. The correct branch is determined by the
limit $G_0(z=z^*=0;\lambda=0)=-\rmi\sigma_x\otimes\openone_2$.

To describe the following steps let us denote the desired solution of
\eqref{eq:pastur} as $G(z,z^*;\alpha,\lambda)$. Now, we proceed as
follows: (i) We determine $G(z_0,z_0^*;0,\lambda)=G_0(z_0,z_0^*;\lambda)$
for a fixed value of $z_0$ (e.g., $z_0=0$) and a finite value of
$\lambda$ (concretely chosen to equal the eventual value of $\alpha$, as
we expect the support of the spectrum to be of that order). (ii) For
values $\alpha$ increasing in small increments from zero to the desired
final value, we solve \eqref{eq:pastur} numerically for
$G(z,z^*;\alpha,\lambda)$, where the initial condition is taken as the
solution from the previous step. (iii) Analogously, we next decrease the
value of $\lambda$ to a small value (here taken as $0.001$; keeping
$\lambda$ small but finite regularizes branch cuts). The same procedure
can be applied to solve \eqref{eq:pastur2}--\eqref{eq:pastur4}.

In practice, we find that  a reliable numerical approximation of the
desired solution $G(z_0,z_0^*;\alpha,0)$ is obtained in a few (about 10)
steps. We can next keep $\alpha$ and $\lambda$ fixed but vary $z$ in
small steps to sample the complex $z$ plane.
 Furthermore, by considering
$z^*$ as a formally independent variable we can also obtain the numerical
derivatives required for the calculation of the eigenvalue probability density
\begin{equation}\label{eq:dosfinal}
\bar\rho(z,z^*)=\frac{1}{2\pi}\frac{\partial}{\partial z^*}(G_{11}+G_{22}).
\end{equation}

\begin{figure}
\hfill{\includegraphics[width=0.8\textwidth]{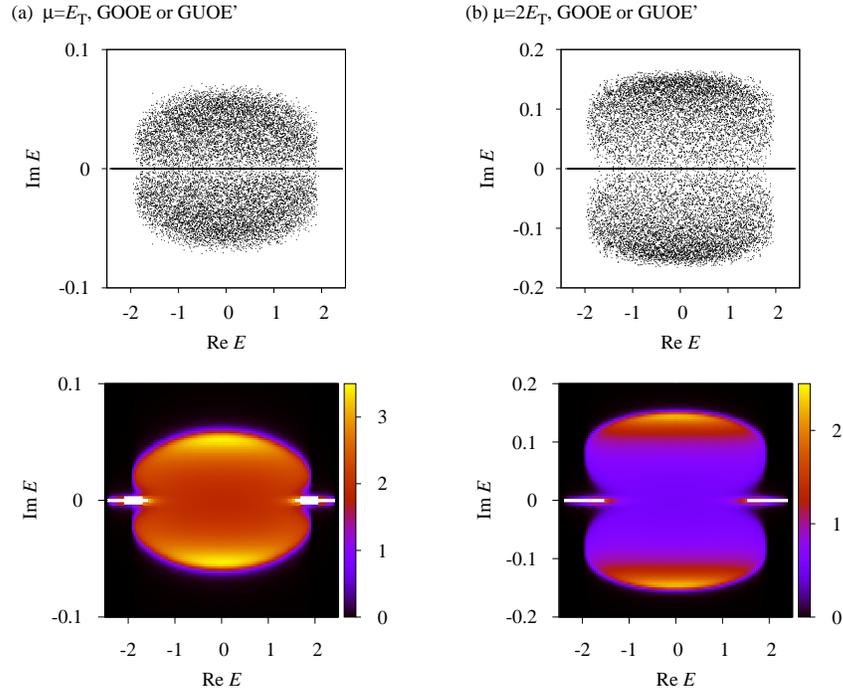}}
\caption{\label{fig:tall}Eigenvalue distributions in the complex plane for the GOOE ensemble
(representing an optical system without magneto-optical effects and uniform amplification or absorption),
 for $N/M=\alpha=0.2$, $T=1$ and (a) $\mu=0.1=E_\mathrm{T}$ as well as
(b) $\mu=0.2=2E_\mathrm{T}$.
The scatter plots in the top panels are
obtained by numerical diagonalization of $20$ matrices $\mathcal{H}$
taken from the GOOE with $M=400$ and $N=80$. The results in the lower panels are obtained from the generalized Pastur
equation \eqref{eq:pastur}, as described in the text. These results also apply to the GUOE$'$.
}
\end{figure}

\begin{figure}
\includegraphics[width=\textwidth]{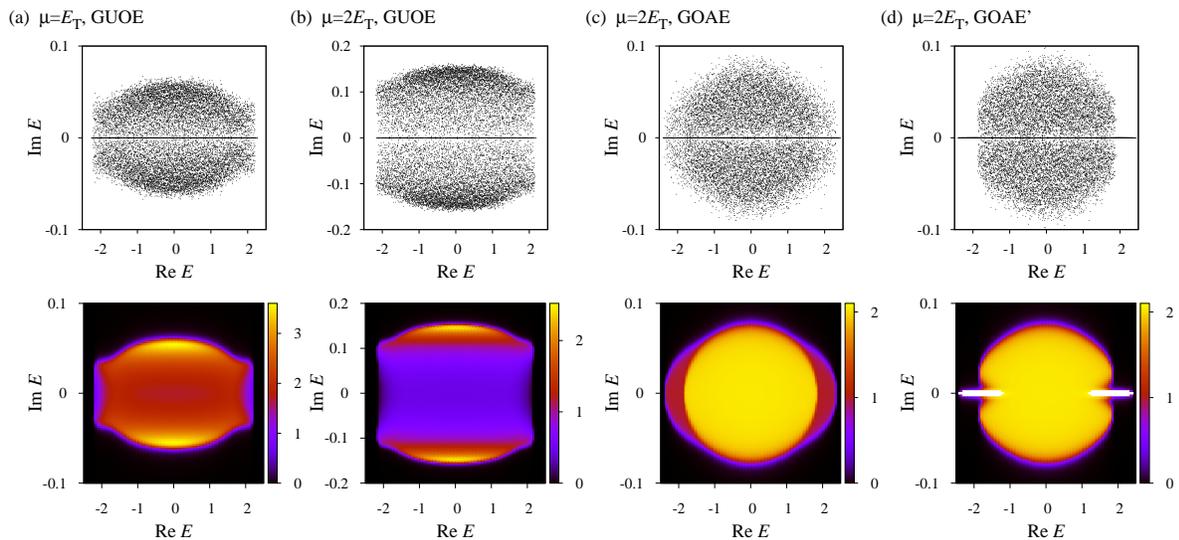}
\caption{\label{fig:tallGUE}(a,b) Same as figure \ref{fig:tall}, but for the GUOE
ensemble. (c,d) Analogous results for the GOAE and  GOAE$'$ ensembles, respectively, with $\mu=2E_\mathrm{T}$.
}
\end{figure}

The lower panels in figure \ref{fig:tall} illustrate the resulting
eigenvalue distribution for the case of the GOOE and GUOE$'$ ensembles,
governed by \eqref{eq:pastur}, as density plots for $\alpha=0.2$ and
$T=1$ ($\gamma=1$), for two different values of $\mu$. The top panels
show the eigenvalues of $20$ random GOOE matrices with $M=400$, $N=80$.
Figure \ref{fig:tallGUE} shows the corresponding results for the GUOE
ensemble, as well as results for the GOAE and GOAE$'$ ensembles for a
single value of $\mu$, obtained from \eqref{eq:pastur2},
\eqref{eq:pastur3} and \eqref{eq:pastur4}, respectively. In all cases
there is excellent agreement, which includes details such as the branches
of the eigenvalue support extending along the real axis in the range
$|\mathrm{Re}\,E|>2$, observed for the GOOE and GOAE$'$ ensembles. (We
also confirmed quantitative agreement by comparing histograms at fixed
$\mathrm{Re}\,E$.)

The one feature which is not captured by the diagrammatic expansion is
the visible accumulation  of eigenvalues in the whole range
$|\mathrm{Re}\,E|<2$ along the real axis. However, we find numerically
that with the present scaling of parameters ($\mu/E_\mathrm{T}$ fixed
independently of $M$), the fraction of these eigenvalues amongst all
eigenvalues steadily decreases $\propto M^{-1/2}$ as $M$ increases, indicating that the real
component of the spectrum indeed becomes negligible in the limit
$M\to\infty$. This is consistent with the earlier prediction for the GOOE
and the GOUE \cite{hsrmt} that the transition to a complex spectrum
happens for a characteristic value $\mu_\mathrm{PT}$ which is much less
than $E_\mathrm{T}$ if $M$ is large. These features render the transition
out of the reach of the described diagrammatic approach. In the following
sections, we will first study the transition in detail based on direct
numerical sampling and diagonalization of the random-matrix ensembles,
and then extend the perturbative treatment of \cite{hsrmt} to quantify
the dependence of $\mu_\mathrm{PT}$ on $T$, $M$ and $N$.

\section{\label{sec:transition}Transition from a real to a complex spectrum}

The transition from  real to  complex-conjugate pairs of eigenvalues can
be quantified by considering the fraction $f_\mathrm{c}$ of eigenvalues
which are complex; $f_\mathrm{c}=0$ indicates a fully real spectrum,
while $f_\mathrm{c}=1$ if the spectrum is fully complex. We determine
this fraction numerically as a function of the non-hermiticity parameter
$\mu$, while keeping $T$, $M$ and $N$ fixed. In the Gaussian ensembles,
the energy levels are taken only from the central region ${\rm
Re}\,E\approx 0$ of the spectrum,  where $\Delta_0\approx \pi/M$ is
approximately constant. This eliminates the spectral edge effects
observed in some of the eigenvalue distributions in the previous
sections, and represents the typical physical conditions met in the
short-wavelength limit of realistic resonators (where the effective level
index is large, and the spectrum is not bounded from above). Furthermore,
for $S_\mathrm{X}=\mathrm{O}$ (uniform absorption and amplification), we
compare the results to the circular ensembles, as in these the mean level
spacing is energy independent. (For $S_\mathrm{X}=\mathrm{A}$ the quantum
maps are less convenient.)

\begin{figure}[t]
  \centering
    \includegraphics[scale=.6]{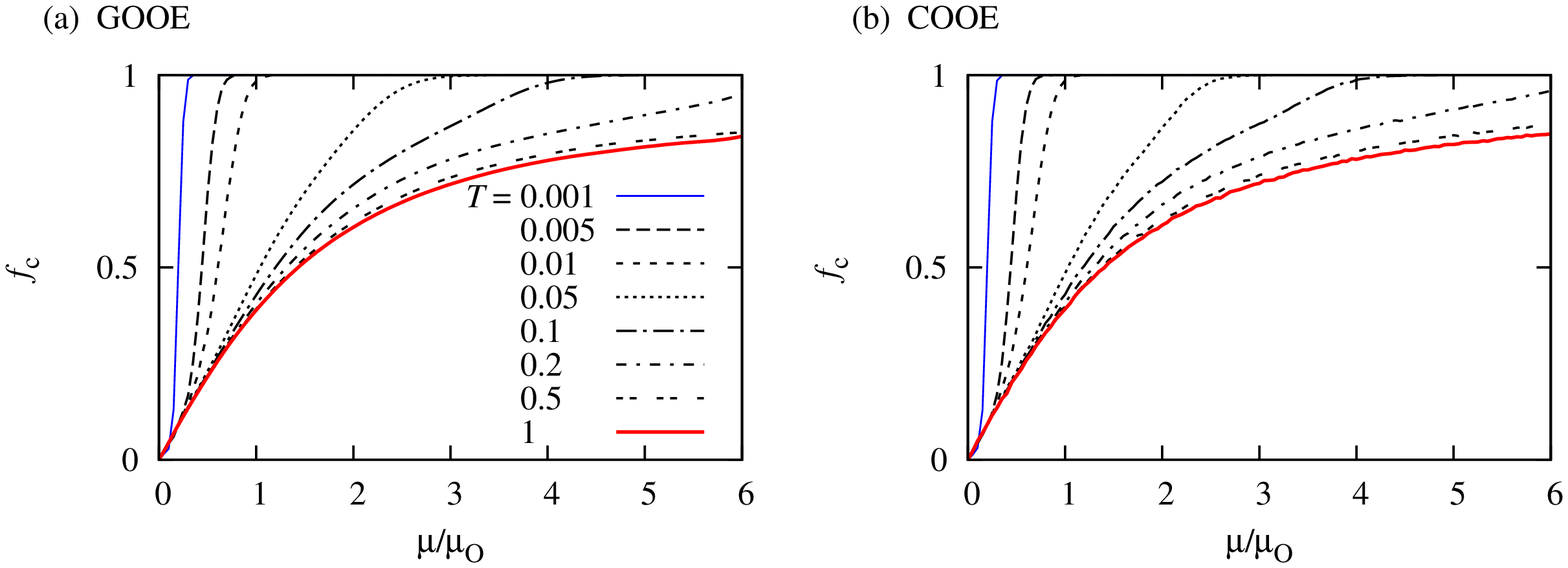}
      \caption{Average fraction $f_\mathrm{c}$ of complex eigenvalues as a function of
the absorption rate $\mu$, scaled to
$\mu_\mathrm{O}=\sqrt{N}\Delta/2\pi$. Panel (a) shows results for the
GOOE around $\mathrm{Re}\,E=0$, while panel (b) shows results for the
COOE, obtained in both cases by numerical sampling of the ensembles with
$M=200$ and $N=40$. The different curves correspond to different
transparencies $T$ of the interface between the amplifying and absorbing
resonators. For $T>T_\mathrm{O}\sim 1/N$, the initial stages of the
transition to a complex spectrum is coupling-independent.}
\label{fig:fc1}
\end{figure}

\begin{figure}[t]
  \centering
    \includegraphics[scale=.6]{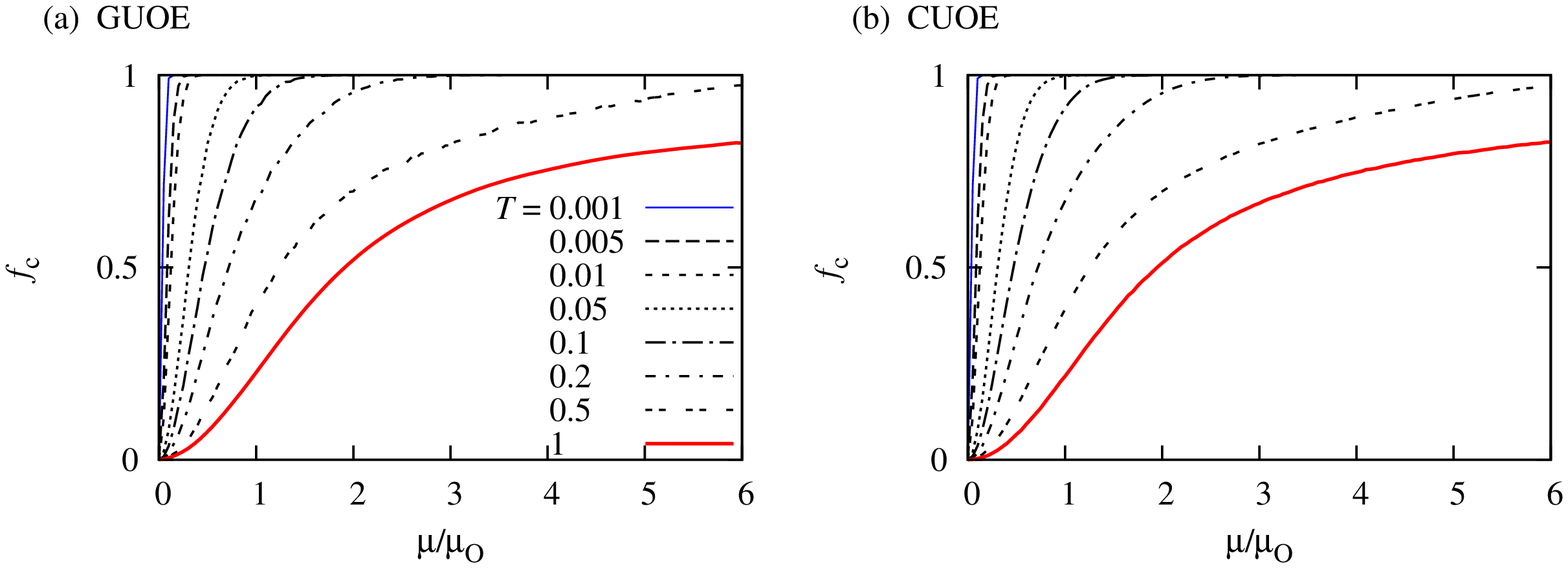}
      \caption{Same as figure \ref{fig:fc1}, but for the GUOE (a) and CUOE (b). Here a
clear dependence on $T$ persists throughout the entire range of this
parameter.} \label{fig:fc2}
\end{figure}

\begin{figure}[t]
  \centering
    \includegraphics[scale=.6]{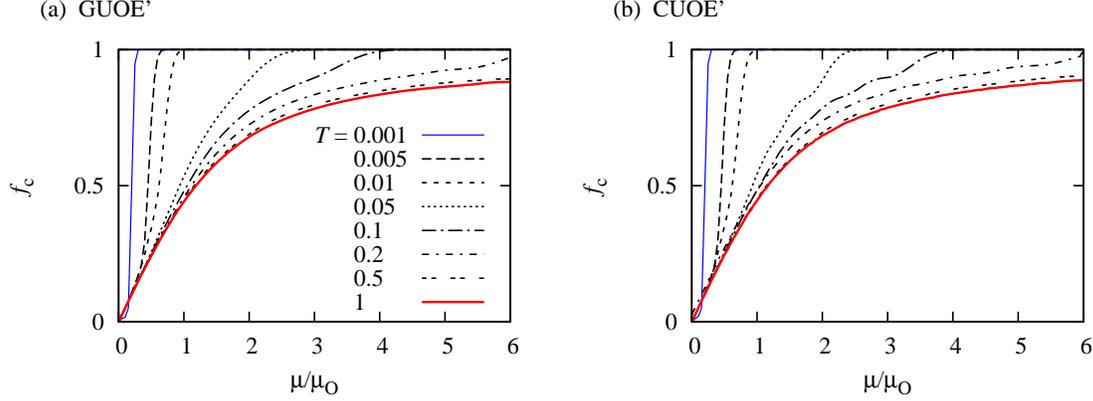}
      \caption{Same as figure \ref{fig:fc1}, but for the GUOE$'$ (left panel) and
CUOE$'$ (right panel). Again, a coupling-independent regime emerges for
$T>T_\mathrm{O}\sim 1/N$.} \label{fig:fc3}
\end{figure}

\begin{figure}[t]
  \centering
    \includegraphics[scale=.6]{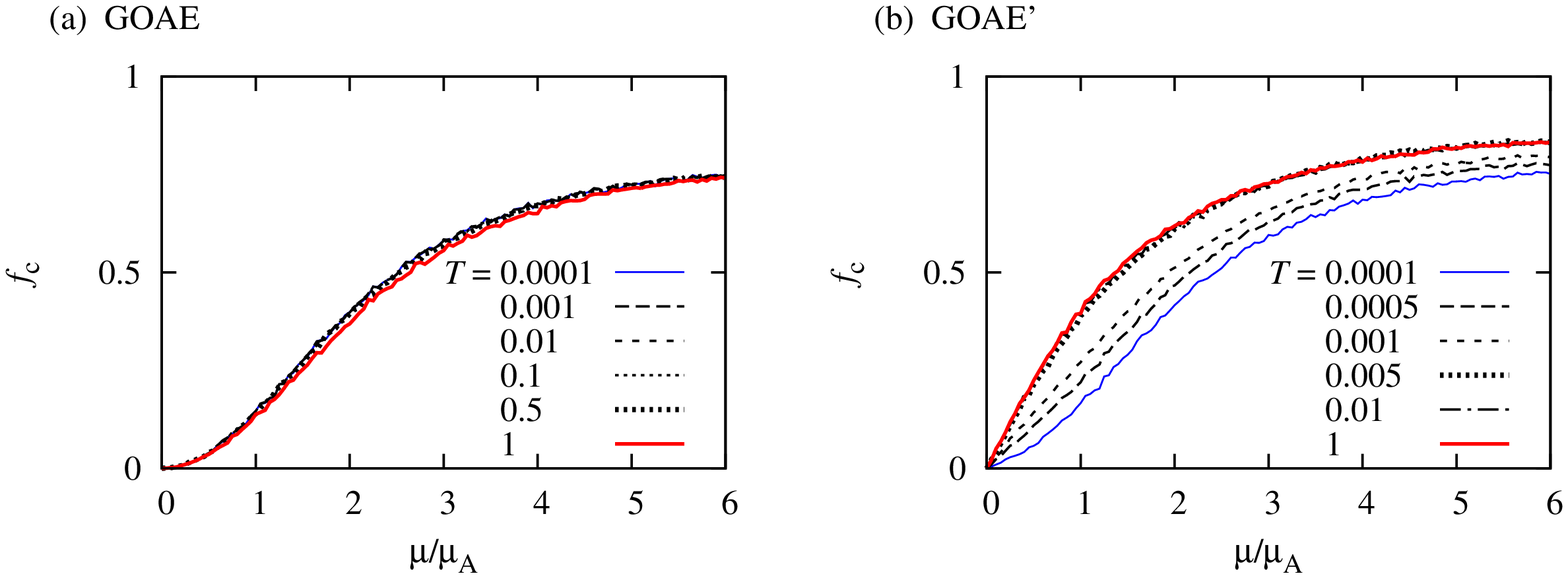}
      \caption{Same as figure \ref{fig:fc1}, but for the GOAE (a) and the GOAE$'$ (b),
and with $\mu$ now given in units of
$\mu_\mathrm{A}=\sqrt{M}\Delta/2\pi$. The transition is almost coupling
independent, with exception of the weak-coupling regime
$T<T_\mathrm{A}\sim 1/N^2$ of the GOAE$'$, where the characteristic scale
$\mu_\mathrm{PT}$ increases as $T\to 0$.} \label{fig:fc4}
\end{figure}

Our eventual goal is to characterize the transition in the different
ensembles by the coupling dependence of the critical scale
\begin{equation}\label{eq:gdef}
\mu_\mathrm{PT}=g(T)\mu_0,
\end{equation}
which we identify via $f_\mathrm{c}(\mu_\mathrm{PT})\sim 1/2$ (without
requiring exact equality). The scale $\mu_0$ is chosen such that the
function $g(T)$ does not depend on $M$ and $N$ if $M\gg N \gg 1$ (with
possible exceptions for weak coupling $T<T_\mathrm{O,A}$, as specified
below). By varying $M$ and $N$ independently we find that this scale
depends on the symmetry $S_\mathrm{X}=\mathrm{O}$ or A, and can be
suitably written as
\begin{equation}\label{eq:mudef}
\mu_0 =\left\{
         \begin{array}{ll}
           \sqrt{N}\Delta/2\pi\equiv \mu_\mathrm{O}, & \hbox{(OO, UO and UO$'$);} \\
           \sqrt{M}\Delta/2\pi\equiv \mu_\mathrm{A}, & \hbox{(OA, OA$'$).}
         \end{array}
       \right.
\end{equation}
This being fixed, we set $M=200$ and $N=40$ and determine
$f_\mathrm{c}(\mu)$ for various values of $T$ with $\mu$ measured in
units of the appropriate $\mu_0$, as shown in figures
\ref{fig:fc1}--\ref{fig:fc4}, and focus the discussion on the
ensemble-specific form of the scaling function $g(T)$ in \eqref{eq:gdef}.

In the GOOE and COOE (figure
\ref{fig:fc1}), we find that
the transition becomes coupling-independent as soon as
$T>T_\mathrm{O}$, where
\begin{equation}
T_\mathrm{O}\sim 1/N.
\end{equation}
In this regime $g(T)\approx 1$; the slight $T$-dependence still observed
in the plots are finite-size effects, which disappear if $M$ and $N$ are
further increased. (However, in this limit $T_\mathrm{O}$ becomes very
small, so that the behaviour for $T<T_\mathrm{O}$ would be difficult to
illustrate; the chosen values of $M$ and $N$ thus constitute a suitable
compromise.) In the GUOE and CUOE (figure \ref{fig:fc2}), on the other
hand, the transition displays coupling dependence throughout the whole
range of $T$.

These results for OO and UO symmetry agree with the predictions in
\cite{hsrmt}, which we systemize in the following section to develop a
microscopic picture that also applies to the remaining ensembles
considered here. The numerical results for these cases are as follows.

In the GUOE$'$ and CUOE$'$ (figure \ref{fig:fc3}), the transition
displays a similar coupling independence as in the GOOE and COOE, with
$\mu_\mathrm{PT}$ (and thus $g(T)$) roughly scaled down by a factor of
about $\sqrt{2}$.

In the GOAE and GOAE$'$ (figure \ref{fig:fc4}), the scale
$\mu_\mathrm{A}$ applies. In both ensembles, there is almost no coupling
dependence throughout the whole range of $T$, with exception of the
weak-coupling regime $T\ll T_\mathrm{A}$ in the GOAE$'$, which is now
delineated by
\begin{equation}
T_\mathrm{A}\sim 1/N^2
.
\end{equation}
Interestingly, in this regime $g(T)$ decreases with increasing $T$, which
amounts to an anomalous behaviour---the coupling between the resonators
enhances the fragility of the real spectrum, in contrast to the usual
situation where increasing the coupling furthers the balance of the
non-hermitian effects in the system.

Note that in the limit $M=N/\alpha\to\infty$ studied in section
\ref{sec:largem}, $\mu_\mathrm{O,A}/E_\mathrm{T}\to 0$ as well as
$T_\mathrm{O,A}\to 0$.

\section{\label{sec:spacings}Relation to level spacing statistics}

To explain the observations of the previous section, we now develop a
microscopic picture of the transition from a real to complex spectrum.
This is based on two ingredients, the level statistics in the hermitian
limit $\mu=0$, and the interaction of the eigenvalues on the real axis as
$\mu$ is increased. Focussing on these two ingredients is motivated by
the fact that the formation of complex eigenvalues requires two real
eigenvalues to coalesce. The required degree of nonhermiticity thus
depends on the distance of the levels at $\mu=0$, and the typical size of
the matrix elements which mix the levels. The ensembles studied here
display different degrees of level repulsion and level mixing, which also
depend on the coupling strength $T$, and our aim is to show that these
characteristics are consistent with the coupling dependence of the
complex fraction $f_\mathrm{c}$ reported in the previous section.

We start with some preliminary observations that justify to separate the
problem of level spacing statistics at $\mu=0$ from the problem of level
mixing at finite $\mu$. First, we note that at $\mu=0$, $T=0$, the system
consists of two uncoupled passive resonators, which both have an
identical real spectrum. In order to inspect how this degeneracy is
lifted, we pass over to a $\mathcal{P}$-symmetric basis,
$\mathcal{H}_{\cal P}=U\mathcal{H}U$, where
$U=2^{-1/2}(\sigma_x+\sigma_z)$ diagonalizes $\sigma_x$, which gives
\begin{eqnarray}
 \mathcal{H}_{\cal P}=
\left(\begin{array}{cc}H+ \Gamma  & -i\mu \\  -i\mu & H-\Gamma\\  \end{array} \right) \quad\mbox{(GOOE and GUOE$'$)}
\label{eq:h1}
\\
 \mathcal{H}_{\cal P}=
\left(\begin{array}{cc}\mathrm{Re}\, H+ \Gamma  & i \mathrm{Im}\, H-i\mu \\  i \mathrm{Im}\,H-i\mu & \mathrm{Re}\,H-\Gamma\\  \end{array} \right) \quad\mbox{(GUOE)}
\label{eq:h2}
\\
 \mathcal{H}_{\cal P}=
\left(\begin{array}{cc}H+A+ \Gamma  & 0 \\  0 & H+A- \Gamma\\
\end{array} \right) \quad\mbox{(GOAE)}
\label{eq:h3}
\\
 \mathcal{H}_{\cal P}=
\left(\begin{array}{cc} H+\Gamma  & A \\  A & H- \Gamma\\  \end{array} \right)
\quad\mbox{(GOAE$'$)}
\label{eq:h4}
\end{eqnarray}

In the hermitian limit $\mu=0$ (implying also $A=0$), all these
transformed Hamiltonians are block diagonal, with exception of
\eqref{eq:h2} for the GUOE. This is the case because in the other cases
$\mathcal{P}$ is an exact symmetry; moreover, for the GOOE, GOAE, and
GOAE$'$, $\mathcal{T}$ and $\mathcal{T}'$ hold separately if $\mu=0$ or
$A=0$. These properties lead to different level statistics in the
hermitian limit, which in all cases but for the GUOE involve the
superposition of two non-interacting level sequences $E^{+}$ and $E^{-}$,
obtained from $H+\Gamma$ and $H-\Gamma$, respectively. The two sequences
are degenerate at $T=0$, but are modified by $\Gamma$, which perturbs the
two sequences in different ways, and because of its positive definiteness
also induces an approximately rigid shift.

In the non-hermitian case (finite $\mu$ or $A$) the $\mathcal{P}$
symmetry remains exact in the GOAE, while $\mathcal{T}'$ symmetry remains
exact in the GOOE and $\mathcal{T}$ symmetry remains exact in the GOAE
and in the GOAE$'$. These differences are reflected in the matrix
elements that mix the level sequences. For this, we recall that in
almost-degenerate perturbation theory, the effective Hamiltonian of two
adjacent levels $E_i$ and $E_j$ is
\begin{equation}
\label{eq:hpert}
\mathcal{H}_2=
\left(\begin{array}{cc} E_i + V_{ii} & V_{ij} \\   V_{ji} & E_j + V_{jj}\\  \end{array} \right),
\end{equation}
where $V$ is a generic perturbation. The perturbation theory is
straightforward at small $T$, but as $T$ increases levels display exact
or avoided crossings. One can then still base estimates by stipulating a
typical spacing $\Delta$ of two adjacent levels at finite $T$, and
restricting the perturbative analysis to the wavefunction overlap
\cite{hsrmt}.

\begin{figure}[t]
  \centering
    \includegraphics[scale=.6]{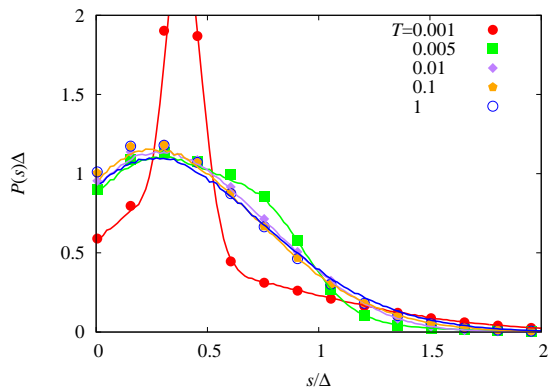}
      \caption{Level spacing distribution of real eigenvalues in the hermitian limit
      of the GOOE (curves) and COOE (points), obtained by numerical sampling of these ensembles with $M=200$ and $N=40$.
      These results also apply to the GOAE$'$ and COAE$'$.}
	\label{fig:ps1}
\end{figure}

\begin{figure}[t]
  \centering
\includegraphics[scale=.4]{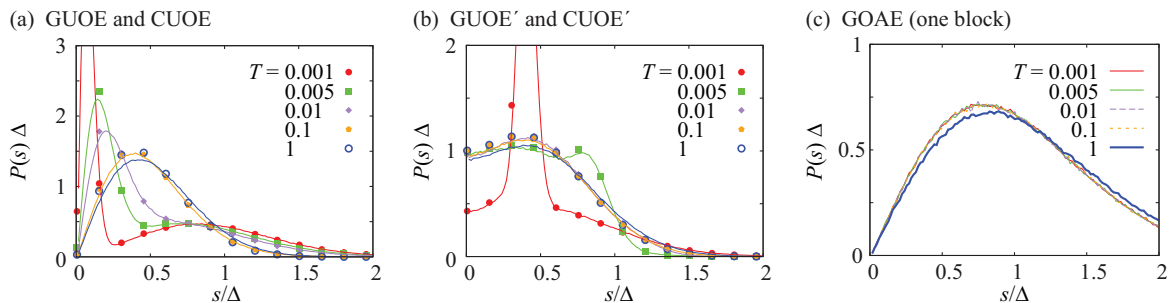}
      \caption{Same as figure \ref{fig:ps1}, but for the GUOE and CUOE (left panel),
 the GUOE$'$ and CUOE$'$ (middle panel), and a single
 block $H+\Gamma$, corresponding to the levels that are mixed by finite $A$ in the GOAE (right panel)}
	\label{fig:ps2}
\end{figure}

As a backdrop for the ensemble-dependent discussion of the details, we
show in figures \ref{fig:ps1} and \ref{fig:ps2} numerically evaluated
level-spacing statistics $P(s)$ at $\mu=0$, where $s$ is the distance
between adjacent levels. In the Gaussian ensembles, we focus again on the
bulk of the spectrum (close $\mathrm{Re}\,E\approx 0$); for $S_\mathrm{X}=\mathrm{O}$
these results are also compared with results from the circular ensembles
(as before $M=200$ and $N=40$).

In figure \ref{fig:ps1}, $P(s)$ is shown for the GOOE and COOE. The data
applies to the full spectrum of $\mathcal{H}$, thus, the superposition of
the sequences $E^+$ and $E^-$ of $\mathcal{H}_\mathcal{P}$ at $\mu=0$
(with mean level spacing $\bar s=\Delta/2$), which is appropriate as
these sequences become mixed by finite $\mu$.
 At small coupling
$T<T_\mathrm{O}\sim 1/N$, the statistics is dominated by the closeness of
levels which degenerate at $T=0$. Based on \eqref{eq:hpert}, one then
finds $\mu_\mathrm{PT}\sim N\sqrt{T}\Delta/2\pi$, thus
$g(T)\sim\sqrt{NT}\ll 1$. For $T>T_\mathrm{O}$, levels in the two
sequences cross (giving $P(0)\approx 1/\Delta$), and the spacing
statistics quickly converges to a coupling-independent form. At finite
$\mu$, adjacent levels of the different sequences $E_n^+$ and $E_m^-$
with $|E_n^+-E_m^-|\sim\Delta$ are mixed by a matrix element of squared
size $|\mu\langle\psi_n^+|\psi_m^-\rangle|^2\sim\mu^2/N$ \cite{hsrmt},
which becomes comparable to $\Delta^2$ at $\mu_{\mathrm{PT}}\sim
\sqrt{N}\Delta/2\pi =\mu_\mathrm{O}$, up to factors of order unity
because of the influence of the level fluctuations. This qualitatively
explains the approximate coupling-independence of $f_\mathrm{c}(\mu)$,
observed for this ensemble in the previous section (figure
\ref{fig:fc1}). One can interpolate between the weak-coupling and
finite-coupling regimes by setting $g(T)\approx (1+1/NT)^{-1/2}$.

The left panel of figure \ref{fig:ps2} shows the analogous result for the
GUOE and CUOE. Here, finite $\Gamma$ also induces a mixing of the
 originally degenerate sequences $E^+$ and $E^-$, which results in a
coupling-dependent level repulsion  (with $P(0)=0$). This corresponds
well to the persistent coupling dependence of $f_\mathrm{c}(\mu)$,
observed in figure \ref{fig:fc2}. Based on \eqref{eq:hpert}, one now
finds $\mu_\mathrm{PT}\sim  \sqrt{NT}\Delta/2\pi$ \cite{hsrmt}, thus
$g(T)\sim\sqrt{T}$, which holds across the whole range of $T$, up to
modifications of order unity, which we now can associate to the influence
of the level statistics.

In the GUOE$'$  and CUOE$'$ (middle panel of figure \ref{fig:ps2}), on
the other hand, levels are again not mixed by finite $T$. Thus, a
coupling-independent level statistics again emerges for $T>T_\mathrm{O}$,
with is similar to the result for the GOOE/COOE (figure \ref{fig:ps1}),
with $P(0)=1/\Delta$. The modal value  is shifted to slightly larger $s$,
in accordance with the larger degree of level repulsion in the standard
GUE/CUE \cite{mehta}. This behaviour corresponds well to the approximate
coupling-independence of $f_\mathrm{c}(\mu)$  in figure \ref{fig:fc3}. We
find in the perturbative treatment that $g(T)$ is the same as in the
GOOE, up to a possible factor of order unity due to the small differences
in the level spacing statistics.

In the GOAE, the superimposed level sequences display the same statistics
as in the GOOE. However, $A$ does not mix these sequences; instead,
eigenvalue coalescence must happen within a given sequence. Therefore, we
consider the spacing within a fixed sequence, which is shown in the right
panel of figure \ref{fig:ps2} (here the mean levels spacing is $\bar
s=\Delta$). The result is almost indistinguishable from the standard
Wigner distribution of the GOE (with $P(s)\propto s$ for small $s$)
\cite{mehta,haake}, as the main effect of $\Gamma$ is a rigid shift;
small deviations appear only for $T\approx 1$. This agrees well the
corresponding behaviour of $f_\mathrm{c}(\mu)$ in figure
\ref{fig:fc4}(a). Perturbatively, the mixing is given by overlaps
$a_{mn}=\langle \psi_n^+|A|\psi_m^+\rangle$ of size
$\overline{a_{mn}^2}=\mu^2/M$, which must be of order $\Delta^2$ for
eigenvalues to coalesce. Thus, we can write
$\mu_\mathrm{PT}\approx\sqrt{M}\Delta/2\pi=\mu_\mathrm{A}$ (up to a
factor of order unity), which is independent of $N$.

The data in  figure \ref{fig:ps1} also applies to the GOAE$'$, which
shows a similar coupling independence, figure \ref{fig:fc4}(b), as the
GOOE and GUOE$'$. The anomalous behaviour at small $T$ can be understood
from the fact that in this regime  eigenvalue coalescence predominantly
occurs between levels $E_n^+$, $E_n^-$ of the two sequences that are
degenerate at $T=A=0$. Thus, perturbatively their eigenvectors
$\psi_n^+\approx\psi_n^-$ are identical, and the first-order coupling
$\langle\psi_n^-|A|\psi_n^+\rangle= 0$ because $A$ is antisymmetric (and
$\psi_n^{\pm}$ is real). For $T>T_\mathrm{A}$, on the other hand, the
coalescence is between originally non-degenerate levels of the two
sequences, and
 $\mu_\mathrm{PT}\sim\sqrt{M}\Delta/2\pi=\mu_\mathrm{A}$ (up
to a factor of order unity, and again independent of $N$), as in the
GOAE.

\section{\label{sec:conclusions}Conclusions}

In summary, motivated by recent optical realizations of non-hermitian
$\mathcal{PT}$-symmetric quantum systems, we identified a number of
symmetry classes which can be realized in optical resonators and differ
by  a choice between two generalized time reversal symmetries
($\mathcal{PT}$ or $\mathcal{PTT}'$), as well as the absence or presence
of magneto-optical effects in the hermitian and nonhermitian parts of the
dynamics. Specifically we considered five scenarios, with symmetries
termed OO, UO, UO$'$, OA and OA$'$.

Our analytical results reveal that in the short-wave limit, the fraction of real
eigenvalues among all eigenvalues in the spectrum decays to zero at any classically finite
amplification and absorption rate $\mu$. Based on numerical results, and
an extension of the perturbative results in \cite{hsrmt}, we find that
the amplification and absorption rate $\mu_\mathrm{PT}\approx \mu_0 g(T)$
at which real eigenvalues turn complex is indeed characterized by a scale
\begin{equation}
\mu_0 =\left\{
         \begin{array}{ll}
           \sqrt{N}\Delta/2\pi\equiv \mu_\mathrm{O}, & \hbox{(OO, UO and UO$'$);} \\
           \sqrt{M}\Delta/2\pi\equiv \mu_\mathrm{A}, & \hbox{(OA, OA$'$),}
         \end{array}
       \right.
\end{equation}
which is classically small but microscopically large. Furthermore, the
scenarios differ in the dependence $g(T)$ on the coupling strength $T$
between the absorbing and amplifying components, which can be explained
in terms of the level spacing statistics in the hermitian limit $\mu=0$,
and distinct mechanisms of how these levels are then mixed when $\mu$ is
finite. For UO symmetry, the transition is $T$-dependent over the whole
range of this parameter. In the OO and UO$'$ classes, a significant dependence is
only observed for $T<T_\mathrm{O} \sim 1/N$, while  the OA$'$ symmetry
class displays an anomalous dependence of the transition on the coupling
strength in the range $T<T_\mathrm{A}\sim 1/N^2$. In the OA symmetry
class, there is negligible coupling dependence over the whole range of
$T$.

The introduced models possess a structure that respects the constraints
imposed by characteristic energy and time scales, the physical nature of
the amplification and absorption, and the accessible coupling strengths
of a realistic (possibly semitransparent) interface. The classification
of these models can be straightforwardly extended to include any
symplectic, chiral, particle-hole like, or additional geometric
symmetries, as previously discussed in hermitian situations
\cite{melsen,zirnbauer,altlandzirnbauer,baranger,whitney1,whitney2}.

In this work we focussed on ensemble-specific spectral properties.
However, as is generally the case in random-matrix theory, there are many
quantities that are less ensemble-specific and should display a large
degree of universality. The prime example is the spectral statistics of
the complex eigenvalues in the bulk of the spectrum. If one is interested
in such statistics, simpler models can be employed. For example, in
scattering theory \cite{fyodorovsommers2} the effective Hamiltonian has a semidefinite
antihermitian part, but the bulk spectral statistics can be studied via
the Ginibre ensemble with complex entries \cite{ginibre}.
Analogously, the spectral constraints of $\mathcal{PT}$ or
$\mathcal{PTT}'$ symmetry are also obeyed by the real Ginibre ensemble,
which has a much simpler matrix structure than our ensembles, and for which
detailed rigorous results are
available \cite{sommers07,forrester,borodin}. That this ensemble is a
good model for other non-hermitian ensembles has been demonstrated, e.g.,
for the case of lattice QCD in \cite{markum}. Notably, in this ensemble,
in the stipulated limit with fixed classical scales, the fraction of real
eigenvalues decays as $1/\sqrt{M}$ \cite{efetov}.

An important constraint in the applicability range of any random-matrix
ensemble is the requirement that many modes are well mixed by multiple
scattering. This is not the case in (quasi) one dimensional
$\mathcal{PT}$-symmetric disordered systems, where states are localized
and the transition happens at much smaller values of $\mu$
\cite{bendix,west,scott}. Furthermore, even in wave-chaotic settings the
multiple scattering can be suppressed by dynamical effects, which can
lead to systematic corrections for the density of  eigenvalues, as
observed in  \cite{birchallweyl} for the strongly amplified states in a
quantum-chaotic system.


\section*{References}

\begin{thebibliography}{98}

\bibitem{bender} Bender, C M and Boettcher, S 1998 Real Spectra in
    Non-Hermitian Hamiltonians Having PT Symmetry \emph{Phys. Rev. Lett.}
    \textbf{80}, 5243--5246 


\bibitem{ptreview1} Bender, C M 2007 Making sense of non-Hermitian
    Hamiltonians \emph{Rep. Prog. Phys.} \textbf{70}, 947--1018

\bibitem{ptreview2} Mostafazadeh, A 2010 Pseudo-Hermitian representation
    of quantum mechanics \emph{Int. J. Geom. Meth. Mod. Phys.}
    \textbf{7}, 1191--1306 

\bibitem{ptexperiments1} Guo, A, Salamo, G J, Duchesne, D, Morandotti, R,
    Volatier-Ravat, M, Aimez, V, Siviloglou, G A and  Christodoulides, D
    N 2009 Observation of PT-Symmetry Breaking in Complex Optical
    Potentials \emph{Phys. Rev. Lett.} \textbf{103}, 093902

\bibitem{ptexperiments2} R{\"u}ter, C E, Makris, K G, El-Ganainy, R,
    Christodoulides, D N, Segev, M and Kip, D 2010 Observation of
    parity-time symmetry in optics \emph{Nature Phys.} \textbf{6},
    192--195 

\bibitem{christodoulidesgroup1} El-Ganainy, R, Makris, K G,
    Christodoulides, D N and  Musslimani, Z H 2007 Theory of coupled
    optical PT-symmetric structures \emph{Opt. Lett.} \textbf{32},
    2632--2634 

\bibitem{christodoulidesgroup2} Makris, K G, El-Ganainy, R,
    Christodoulides, D N and Musslimani,  Z H 2008 Beam Dynamics in PT
    Symmetric Optical Lattices \emph{Phys. Rev. Lett.} \textbf{100},
    103904 



\bibitem{christodoulidesgroup3} Musslimani, Z H, Makris, K G, El-Ganainy,
    R and Christodoulides, D N 2008 Optical Solitons in PT Periodic
    Potentials \emph{Phys. Rev. Lett.} \textbf{100}, 030402

\bibitem{pteffects1} Longhi, S 2009 Bloch Oscillations in Complex
    Crystals with PT Symmetry \emph{Phys. Rev. Lett.} \textbf{103},
    123601 



\bibitem{pteffects2a} Ramezani, H,  Kottos, T, El-Ganainy, R and
    Christodoulides, D N 2010 Unidirectional nonlinear PT-symmetric
    optical structures \emph{Phys. Rev. A} \textbf{82}, 043803


\bibitem{pteffects4} Longhi, S 2011 Invisibility in PT-symmetric complex
    crystals \emph{J. Phys. A: Math. Theor.} \textbf{44}, 485302

\bibitem{pteffects5} Lin, Z,  Ramezani, H, Eichelkraut, T, Kottos, T,
    Cao, H and Christodoulides, D N 2011 Unidirectional Invisibility
    Induced by PT-Symmetric Periodic Structures \emph{Phys. Rev. Lett.}
    \textbf{106}, 213901 

\bibitem{hsqoptics} Schomerus, H  2010 Quantum Noise and Self-Sustained
    Radiation of PT-Symmetric Systems \emph{Phys. Rev. Lett.}
    \textbf{104}, 233601 

\bibitem{longhilaserabsorber} Longhi, S  2010 PT-symmetric laser absorber
    \emph{Phys. Rev. A} \textbf{82}, 031801(R)

\bibitem{variousstonepapers1} Chong, Y D, Ge, L and  Stone, A D 2011
    PT-Symmetry Breaking and Laser-Absorber Modes in Optical Scattering
    Systems \emph{Phys. Rev. Lett.} \textbf{106}, 093902


\bibitem{yoo} Yoo, G, Sim, H-S  and Schomerus,  H 2011 Quantum noise and
    mode nonorthogonality in non-Hermitian PT-symmetric optical
    resonators \emph{Phys. Rev. A} \textbf{84}, 063833

\bibitem{mehta} Mehta,  M L 2004 \emph{Random Matrices}, 3rd ed New York,
    NY: Elsevier

\bibitem{haake} Haake,  F  2010 \emph{Quantum signatures of chaos}, 3rd
    ed Berlin: Springer

\bibitem{hsrmt} Schomerus, H 2011 Universal routes to spontaneous
    PT-symmetry breaking in non-Hermitian quantum systems \emph{Phys.
    Rev. A} \textbf{83}, 030101(R) 

\bibitem{birchallweyl} Birchall, C and Schomerus, H 2012 Fractal Weyl
    laws for amplified states in PT-symmetric resonators arXiv:1208.2259

\bibitem{hsreview} Schomerus, H 2012 From scattering theory to complex
    wave dynamics in non-hermitian PT-symmetric resonators
    arXiv:1207.1454


\bibitem{magnea} Magnea, U 2008 Random matrices beyond the Cartan
    classification \emph{J. Phys. A: Math. Theor.} {\bf 41}, 045203

\bibitem{akemann1} Akemann, G, Damgaard, P H, Osborn J C and Splittorff,
    K 2007 A new Chiral Two-Matrix Theory for Dirac Spectra with
    Imaginary Chemical Potential \emph{Nucl. Phys.  B} {\bf 766} 34--67


\bibitem{akemann2} Akemann, G 2011 Non-Hermitian extensions of Wishart
    random matrix ensembles \emph{Acta Physica Polonica B} \textbf{42}
    901--921 

\bibitem{melsen} Melsen, J A, Brouwer, P W, Frahm, K M and Beenakker, C W
    J 1996 Induced superconductivity distinguishes chaotic from
    integrable billiards \emph{Europhys. Lett.} \textbf{35}, 7.

\bibitem{zirnbauer} Zirnbauer, M 1996 Riemannian symmetric
    superspaces and their origin in random-matrix theory
    \emph{J. Math. Phys.} \textbf{37}, 4986--5018

\bibitem{altlandzirnbauer} Altland, A and Zirnbauer, M R 1997 Nonstandard
    symmetry classes in mesoscopic normal-superconducting hybrid
    structures \emph{Phys. Rev. B} \textbf{55}, 1142--1161

\bibitem{stoffregen} Stoffregen, U, Stein, J,  St{\"o}ckmann, H-J,
    Ku{\'s}, M and Haake, F 1995
Microwave Billiards with Broken Time Reversal Symmetry
    \emph{Phys. Rev. Lett.} \textbf{74}, 2666.

\bibitem{janik1} Janik, R A, Nowak, M A, Papp, G, Wambach, J and Zahed, I
    1997
    Non-Hermitian random matrix models: Free random variable approach
    \emph{Phys. Rev. E} \textbf{55}, 4100--4106

\bibitem{janik2} Janik, R A, Nowak, M A, Papp, G, and Zahed, I 1997
    Non-hermitian random matrix models \emph{Nucl. Phys. B} \textbf{501}, 603--642

\bibitem{fyodorovsommers2} Fyodorov, Y V and Sommers, H-J 2003 Random
    matrices close to Hermitian or unitary: overview of methods and
    results \emph{J. Phys. A: Math. Gen.} \textbf{36} 3303--3347

\bibitem{baranger} Baranger, H U and Mello, P A 1996 Reflection symmetric
    ballistic microstructures: Quantum transport properties \emph{Phys.
    Rev. B} \textbf{54}, R14297 

\bibitem{whitney1} Whitney, R S, Schomerus, H and Kopp, M 2009
    Semiclassical transport in nearly symmetric quantum dots. I. Symmetry
    breaking in the dot
    \emph{Phys. Rev. E} \textbf{80}, 056209

\bibitem{whitney2} Whitney, R S, Schomerus, H and Kopp, M 2009
    Semiclassical transport in nearly symmetric quantum dots. II. Symmetry breaking due to asymmetric leads
    \emph{Phys. Rev. E} \textbf{80}, 056210


\bibitem{ginibre} Ginibre, J 1965
    Statistical Ensembles of Complex, Quaternion, and Real Matrices
    \emph{J. Math. Phys.} {\bf 6}, 440--449

\bibitem{sommers07}  Sommers, H-J 2007
    Symplectic structure of the real Ginibre ensemble
    \emph{J. Phys. A: Math. Theor.} \textbf{40}, F671--F676

\bibitem{forrester}   Forrester, P J and Nagao, T 2007
    Eigenvalue Statistics of the Real Ginibre Ensemble \emph{Phys. Rev. Lett.}
    \textbf{99}  050603

\bibitem{borodin}
    Borodin, A and Sinclair, C D, 2007
    Correlation Functions of Ensembles of
Asymmetric Real Matrices  arXiv:0706.2670v2

\bibitem{markum} Markum, H, Pullirsch, R and Wettig, T 1999
    Non-Hermitian Random Matrix Theory and Lattice QCD with Chemical Potential
    \emph{Phys. Rev. Lett.} \textbf{83}, 484--487

\bibitem{efetov} Efetov, K B 1997 Directed Quantum Chaos \emph{Phys. Rev.
    Lett.} {\bf 79}, 491

\bibitem{bendix} Bendix, O, Fleishmann, R, Kottos, T and Shapiro,  B 2009
    Exponentially Fragile PT Symmetry in Lattices with Localized
    Eigenmodes \emph{Phys. Rev. Lett.} \textbf{103}, 030402

\bibitem{west} West, C T, Kottos, T and Prosen,  T 2010 PT-Symmetric Wave
    Chaos \emph{Phys. Rev. Lett.} \textbf{104}, 054102

\bibitem{scott} Scott, D D and Joglekar, Y N 2011 Degrees and
    signatures of broken PT symmetry in nonuniform lattices
    \emph{Phys. Rev. A} \textbf{83}, 050102(R)


\end{thebibliography}
\end{document}